\documentclass[10pt]{emulateapj}
\newcommand{\beq}{\begin{equation}}
\newcommand{\eeq}{\end{equation}}
\newcommand{\bea}{\begin{array}}
\newcommand{\eea}{\end{array}}

\shorttitle{Mean motion resonance formation in the planetary systems}
\shortauthors{Wang \& Ji}

\begin{document}

\title{Near Mean-motion Resonances in the Systems Observed by Kepler: Affected by Mass Accretion and Type I Migration}

\author{Su Wang\altaffilmark{1} and Jianghui Ji\altaffilmark{1}}
\altaffiltext{1}{CAS Key Laboratory of Planetary Sciences, Purple
Mountain Observatory, Chinese Academy of Sciences, Nanjing 210008,
China; wangsu@pmo.ac.cn, jijh@pmo.ac.cn.}

\begin{abstract}

The Kepler mission has released over 4496 planetary candidates, among which 3483 planets have been confirmed as of April 2017. The statistical results of the planets show that there are two peaks around 1.5 and 2.0 in the distribution of orbital period ratios. The observations indicate that a plenty of planet pairs could have firstly been captured into mean motion resonances (MMRs) in   planetary formation. Subsequently, these planets depart from exact resonant locations to be near MMRs configurations.  Through type I migration, two low-mass planets have a tendency to be trapped into first-order MMRs (2:1 or 3:2 MMRs), however two scenarios of mass accretion of planets and potential outward migration play an important role in reshaping their final orbital configurations. Under the scenario of mass accretion, the planet pairs can cross 2:1 MMRs and then enter into 3:2 MMRs, especially for the inner pairs. With such formation scenario, the possibility that two planets are locked into 3:2 MMRs can increase if they are formed in a flat disk. Moreover, the outward migration can make planets have a high likelihood to be trapped into 3:2 MMRs.  We perform additional runs to investigate the mass relationship for those planets in three-planet systems, and we show that two peaks near 1.5 and 2.0 for the period ratios of two planets can be easily reproduced through our formation scenario.  We further show that the systems in chain resonances (e.g., 4:2:1, 3:2:1, 6:3:2 and 9:6:4 MMRs), have been observed in our simulations. This mechanism can be applicable to understand the formation of systems of Kepler-48, Kepler-53, Kepler-100, Kepler-192, Kepler-297, Kepler-399, and Kepler-450.

\end{abstract}


\section{Introduction}
The Kepler Space Telescope has monitored more than 150,000 stars for four years in an aim to discover Earth-like planets. As of April 2017, there are over 4496 planetary candidates and 3483 of them have been confirmed to be exoplanets, based on the released data \citep{Bata13, Mazeh13, Fab14}. The discovery of Kepler mission provides us an abundant sample of planetary systems to understand their formation and evolution \citep{MB16, GJ17, MF17}. Among them, there are $\sim$ 581 multiple planetary systems. The architectures of the systems further supply us with lots of clues hidden in the formation process \citep{MS14}. From a statistical analysis of Kepler data, we show that in the Figure 1 of \citet{Wang14} there are a great number of adjacent planet pairs locating at near 1.5 and 2.0 in the distribution of orbital period ratios \citep{Lissauer11,lee13}, indicating that the planet pairs may be firstly trapped into the first-order MMRs, and then form the near MMRs configuration. The released data reports that 20.5\% and 10.5\% of planet pairs are in the near 2:1 or 3:2 MMRs, respectively. Our earlier studies show that 18.0\% and 7.0\% of planet pairs in three-planet systems are in near 2:1 or 3:2 MMRs \citep{Wang12, Wang14}. These kinds of MMRs suggest the existence of a so-called chain resonances, such as 4:2:1, 3:2:1 and 6:3:2 MMRs. Moreover,  K2, as a follow-up program of Kepler mission, has started to search for transiting planets within 100 sq deg fields \citep{Dr17} and has already reported $\sim$ 197 planet candidates \citep{Cre15, Van15, Cre16}. From K2 data, it is common that the planet pairs of multiple planetary systems move near MMRs. For example, K2-72, hosting four terrestial planets with masses of 1.1 to 2.6 $m_\oplus$, bears near MMRs configuration \citep{WM14, Cre16, Dr17}, whereas K2-19 consisting of two planets is approximate to 3:2 MMR configuration \citep{Ar15}. Besides, the system TRAPPIST-1 was discovered to have seven planets in chain MMRs, where two pairs of them are in 3:2 MMRs \citep{Gi17}. Kepler-60, which bears three planets in 5:4:3 MMRs, is in the generalized Laplace resonance \citep{goz16}. Moreover, the planet pairs with Jupiter-mass are found to be near MMRs \citep{lee13}. For example, three Jovian planets in the system GJ 876 are reported to be in Laplacian resonance configuration, where two pairs are both in 2:1 MMRs \citep{Marcy01, Rivera10, marti13, bat15, nelson16, marti16}. Therefore, one of the motivations for this work is to investigate the near MMRs in the planetary systems, in an attempt to understand the formation and evolution of these systems.

The convergent migration scenario is the most classical theory to explain the formation of these orbital configurations \citep{GT80, lin96,Bryden00,MS01, miga16, Liu17a, Liu17b, iz17, ramos17}. For example, the 2:1 MMRs in GJ 876 system can be formed considering the inward migration scenario \citep{lee02, Ji02, Ji03, Zhou05, Zhang10, nelson16}. Two well-separated equal-mass planets can be captured into a first-order MMR from numerical simulations by \cite{Ogihara13}. Even in our solar system, the Kuiper Belt Objects (KBOs) which are in 3:2 or 2:1 MMRs with Neptune can be explained by the migration of Neptune \citep{Malhotra95}.

In our earlier investigations \citep{Wang12, Wang14, Sun17}, we presented the 4:2:1 MMRs formation of KOI-152 system with three planets. In the work, we assumed that the planets are formed with their nominal masses in the outer region of the system, subsequently they undergo orbital migration due to mutual interaction with gas disk. In the migration, three planets are found to be trapped into first-order MMRs. Finally, tidal effect from the central star may drive the planets move to near MMRs region. In this formation scenario, three planets suffer type I migration, and halt at the location of density maximum. Considering this formation scenario, we further investigated how the final configurations had been affected by star magnetic field, star accretion rate, speed of type I migration. We showed that the possibility of two planets trapped into 2:1 MMRs are very high and the 4:2:1 MMRs seems to be common in the planetary systems. However, from the numerical simulations, we observe that the formation of 3:2 MMRs is not so easy to reach as compared with the 3:2 MMRs cases from Kepler data.

As a matter of fact, there are two important mechanisms of the planetary mass accretion scenario in various profiles of disks and the potential outward type I migration, which will play part in the final configurations of the systems. Herein we will summarize several key points as follows.

\begin{enumerate}
\item
First of all, the mass accretion process will alter the masses of planets during the evolution of systems. The final mass distribution of planets in the system is helpful to justify which kind of mass relationship is crucial in the formation of near MMRs. Secondly, the change of mass will alter the gravitational influence on other planets which may affect the final configuration of the system. Finally, the migration timescale is inversely proportional to the mass of planets, and the speed of type I migration is sensitive to final configuration \citep{Wang14}. Therefore, the mass growing process is a considerable factor to be explored in the formation of near MMRs. In the work of \cite{Pet13}, they took into account the mass accretion  in the configuration formation, and found that the mass growing plays a vital role of understanding the formation of 3:2 MMRs.  But in their work, they simply changed the planetary masses in situ in the range of 20-100 $M_\oplus$ rapidly. However, the distribution of final masses , which may have influence on final configuration of systems,  will be significantly different due to the existence of orbital migration and various profile of disks.

\item
Bulk densities of several super-Earths have been obtained with the combination of transiting and radial velocity \citep{Lxw12, Mar14}. Comparing with the pure rocky composition, super-Earths are suggested to hold a hydrogen-helium atmosphere due to their low bulk density \citep{Jin14,Owen16,Jin17}. This indicates that most of them have been formed before the depletion of gas disk \citep{LF14, Ro15}. As shown in the work of \citet{Wang12}, \citet{Wang14}, and \citet{ Sun17}, for the formation of terrestrial planet with a surrounding gaseous disk, the type I orbital migration is an inevitable process due to mutual interaction between planets and gaseous disk \citep{idalin04}. With the estimated linear model, the fast inward migration will lead to a high possibility of planet pairs in 2:1 MMRs \citep{Wang12, Wang14}. According to hydrodynamical simulations, the speed of type I migration can be slowed down or even be reversed \citep{KC08, Kley09, Wang11, Bans15, Ogihara15, Uri15}. In our previous study, we have investigated the role of speed of type I migration. But the effect of potential outward migration is still not clear. The outward migration not only results in diverse separations between two planets \citep{Liu15}, but also alters the final location of the planets which will cause the variations of final masses. Therefore, the outward orbital migration can act as an alternative scenario that may change the final configuration of planetary systems significantly.
\end{enumerate}

In this work, we mainly focus on exploring the formation of near mean motion resonances affected by the mass accretion process of the planets and outward orbital migration in the system. There are mainly four factors that will influence the final configuration of the system: (1) the density profile of the solid disk which determines the speed of mass accretion of planets, (2) the density profile of the gas disk which affects the speed of type I orbital migration and the final positions of planets, (3) the initial masses of planets that affect the mass accretion and the final masses of the planets, and (4) the direction of orbital migration. In Section 2, we introduce our models including the gas and solid disk models, mass accretion process, the orbital migration and gas damping scenario. in Section 3, we present the major results of six Groups by considering various initial planetary masses and migration modes from our numerical simulations. Section 4 shows the main conclusions and discussions.

\section{Models}
\subsection{Disk Models and Mass Accretion Process}
The surface density profile of solid  disk ($\Sigma_{\rm d}$) based on the empirical minimum-mass solar nebular model (MMSN; Hayashi 1981) at a stellar distance $a$ are described as
\begin{equation}
\Sigma_{\rm d}=10f_{\rm d}\eta_{\rm ice}(\frac{a}{\rm {1AU}})^{-s}\rm{gcm^{-2}}, \label{denso}
\end{equation}
where $f_{\rm d}$ is the enhancement factor of the MMSN, $s$ is the power low index of the solid density  and $\eta_{\rm ice}$ is the volatile enhancement factor, respectively. Herein $\eta_{\rm {ice}}=1$ is interior to the snow line, whereas $\eta_{\rm {ice}}=4.2$ is exterior to the snow line. The location of the snow line $a_{\rm {ice}}$ is 2.7 AU for the system with a sun-like star.  According to the Kepler data, the mass of planets can reach tens of Earth-mass. Thus, in order to yield large mass planets, the solid disk may be massive than the MMSN. Herein we set $f_d=3$, the solid material in the disk can reach 10 to 55 Earth-mass in the range of [0.05, 2] AU depending on the value of $s$. The surface density profile of the gas disk is denoted as
\begin{equation}
\Sigma_{\rm g}=\frac{\dot M}{3\pi \alpha (a) c_s h}{\rm
exp}\left(\frac{-t}{\tau_{\rm dep}}\right)\eta,\label{dens}
\end{equation}
where $t$ means the time and $\tau_{\rm dep}$ denotes the gas disk depletion timescale which is estimated to be a few million years \citep{hai01}. The density scale height $h=c_s/\Omega$, $c_s=\sqrt{kT(r)/(\mu m_{pro})}$ is the speed of sound at the mid-plane, where $k$ is the Boltzmann constant, $m_{pro}$ is the mass of proton, and $T(r)\propto r^{-q}$ is the temperature at the mid-plane. The star accretion rate $\dot M$ can be evaluated as \citep{natta,Vor09}
\begin{equation}
\dot M\simeq 2.5\times
10^{-8}\left(\frac{M_*}{M_\odot}\right)^{1.3\pm 0.3}M_\odot~ {\rm
yr}^{-1}. \label{mdot}
\end{equation}

Herein we adopt $\dot M=1.0\times 10^{-9}$$M_\odot~ \rm {yr^{-1}}$.
Additionally, $h$ is the disk scale height, $\alpha$ is the efficiency factor of angular
momentum transport.  $\alpha$ and $\eta$ are expresses as
\begin{equation}
\alpha_{\rm eff} (a)=\frac{\alpha_{\rm dead}-\alpha_{\rm mri}}{2}
\left[{\rm erf}\left(\frac{a-a_{\rm crit}}{0.1 a_{\rm
crit}}\right)+1\right]+\alpha_{\rm mri},
 \label{alpeff}
\end{equation}
\begin {equation}
\eta=0.5\left[{\rm erf}\left(\frac{a-a_{\rm mstr}}{0.1a_{\rm
mstr}}\right)+1\right],
\end{equation}
where $a_{\rm crit}$ represents the location of the boundary of MRI and $a_{\rm mstr}$ stands for the location of truncation of the magnetic field. They are modeled as
\citep{konigl,KL07,KL09}

\begin{eqnarray}
a_{\rm crit}=0.16 ~{\rm
AU}~\left(\frac{\dot{M}}{10^{-8}M_{\odot}~{\rm
yr}^{-1}}\right)^{4/9} \left(\frac{M_{*}}{M_{\odot}}\right)^{1/3}
\nonumber\\
\times\left(\frac{\alpha_{\rm
mri}}{0.02}\right)^{-1/5}\left(\frac{\kappa_D}{1{\rm
cm^2g^{-1}}}\right),~~~~
 \label{acrit}
\end{eqnarray}

and

\begin{eqnarray}
a_{\rm mstr}=(1.06\times 10^{-2} ~{\rm AU}) \beta'
\left(\frac{R_*}{R_\odot}\right)^{12/7}\left(\frac{B_*}{1000{\rm
G}}\right)^{4/7}
\nonumber\\
\times\left(\frac{M_*}{M_\odot}\right)^{-1/7}\left(\frac{\dot{M}}{10^{-7}M_\odot~
{\rm yr^{-1}}}\right)^{-2/7},~~ \label{mstr}
\end{eqnarray}
where $\kappa_D$ is the grain opacity, we choose $\kappa_D=0.2$ in this work. $B_*$ is the star magnetic field. We adopt $B_*=2.5$ KG for a star with relative high magnetic field. $\beta'=1$ represents a typical Alfv\'{e}n radius in a spherical accretion mode. $\alpha_{\rm dead}=0.001$
and $\alpha_{\rm mri}=0.01$ represent the value at the mid-plane of the disk in the dead zone and active zone, respectively.

Considering the above formulas, the surface density profile of the gas disk can be expressed as $\Sigma_g\propto r^{(-3/2+q)}$. Herein, we set $(q=3/2-s)$ to make $\Sigma_g\propto r^{(-s)}$ which is consistent with the profile of solid disk.

\begin{figure}[ht!]

  \epsscale{1.2}
  \plotone{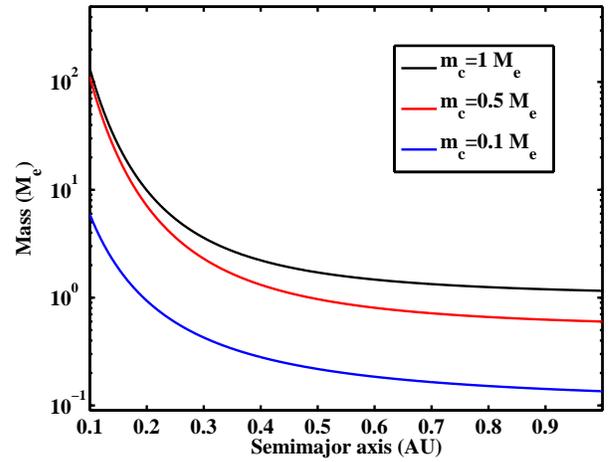}
 \caption{The mass accretion process based on equation (\ref{massacc}) with different initial core masses. The black, red, and blue lines represent the planetary core with 1, 0,5, and 0.1 $m_\oplus$, respectively. They migrate from 1 AU initially. The boundary of the inner hole of the disk locates at 0.1 AU in this case.
 \label{mass}}
\end{figure}

Under such solid and gas disk profiles, the mass accretion of planetesimals with a core mass $m_c$ can be described as  \citep{idalin04}
\begin{equation}
\dot M_c=2.26\times10^{-7}\times (\frac{a}{1AU})^{-(0.5+s)}(\frac{m_c}{M_\oplus})^{2/3}M_{\oplus}\rm yr^{-1}. \label{massacc}
\end{equation}
In this work, the planetary embryos can grow up following equation (\ref{massacc}). When they enter into the inner hole of the disk, the mass accretion process will halt because of the lack of materials. Figure \ref{mass} shows an estimated growing process with various initial core masses.

\subsection{Eccentricity damping and planetary migration }
During the formation process, the planetary embryos are surrounded by the gas, eccentricity damping will be produced by the mutual interactions between embryos and gas. The eccentricity $e$ will be damped in a timescale $\tau_{\rm damp}$ described as \citep{Cre06}

\begin{eqnarray}
\tau_{\rm damp}=\left(\frac{e}{\dot
e}\right)=\frac{Q_e}{0.78}\left(\frac{M_*}{m}\right)
\left(\frac{M_*}{a^2\Sigma_g}\right)\left(\frac{h}{r}\right)^4\Omega^{-1}
\nonumber\\
\times\left[1+\frac{1}{4}\left(e\frac{r}{h}\right)^3\right] {\rm
yr},~~~~~~~~~~~~~~~~~ \label{damp}
\end{eqnarray}
where $Q_e$ is a normalized factor. We choose $Q_e=0.1$ in the simulations to fit with hydrodynamical
simulation results. ($h/r$) represents the ratio between disk scale
height and distance from the central star. $\Omega$ means the Kepler angular velocity.

Additionally, the orbital migration of planets which is triggered by the angular momentum exchange between gas disk and embedded planets \citep{idalin04}. Using the analyzed linear model on isothermal gaseous disk, the timescale of type I migration can be described as \citep{GT79,Ward97,Tan02}

\begin{eqnarray}
\tau_{\rm migI}=\frac{a}{|\dot{a}|}=\frac{1}{f_1(2.7+1.1s)}
\left(\frac{M_*}{m}\right)\left(\frac{M_*}{\Sigma_ga^2}\right)
\nonumber\\
\times\left(\frac{h}{a}\right)^2
\left[\frac{1+(\frac{er}{1.3h})^5}{1-(\frac{er}{1.1h})^4}\right]\Omega^{-1}\rm{yr},~~~~~~
\label{tauI}
\end{eqnarray}
where $f_1$ is the reduction factor \citep{Wang12, Wang14}. In this work, we assume $f_1=0.3$ to obtain a system which hosts three planets approximate to 4:2:1 MMRs in the model case. Using Equation \ref{tauI}, the direction of type I orbital migration is inward. And planets can stop migrating at the boundary of the inner hole or the boundary of MRI \citep{Wang12, Wang14}. The outward migration will be discussed in Section 3.3.

\begin{figure*}[ht!]
\flushleft

  \epsscale{1.15}\plotone{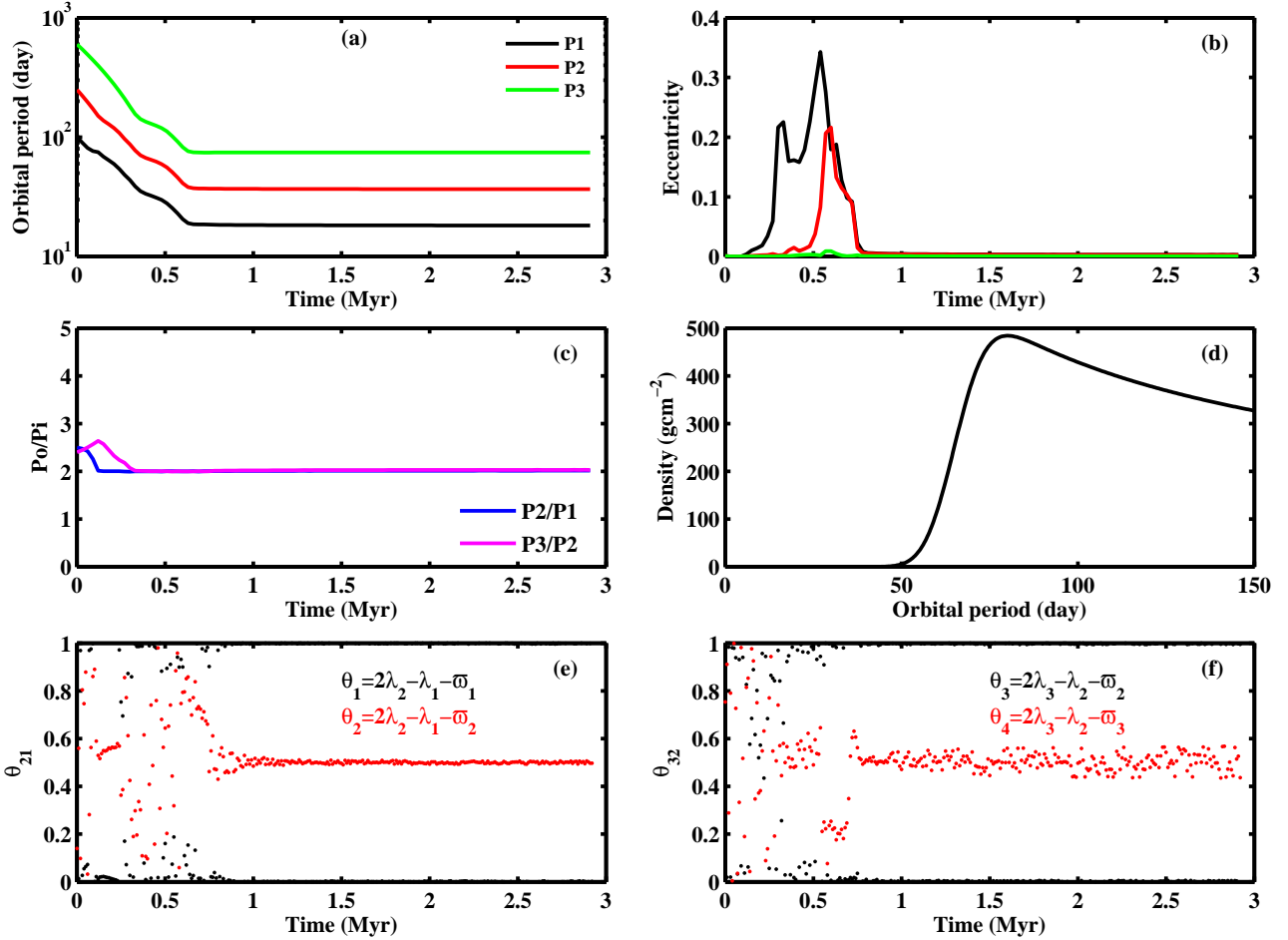}
 \caption{The model case that three planets trapped into 4:2:1 MMRs. The envoluiton of orbital period, eccentricity, period ratio, and the resonance angles are shown in panel (a), (b), (c), (e), and (f). The gas density profile is shown in panel (d). In panel (a) and (b), the black line means the planet 1, red line represents planet 2, and green line displays planet 3. In panel (c), the purple line shows the evolution of period ratio between planet 3 and planet 2, while the blue line means that of planet 2 and planet 1.
 \label{modelc}}
\end{figure*}

Beside the orbital migration and the gas damping effect, we also consider the gravitational interaction between the planetary embryos and the central star. The acceleration of the planetary embryos with mass $m_i$ is described as

\begin{eqnarray}
\frac{d}{dt}\textbf{V}_i =
 -\frac{G(M_*+m_i
)}{{r_i}^2}\left(\frac{\textbf{r}_i}{r_i}\right) +\sum _{j\neq i}^N
Gm_j \left[\frac{(\textbf{r}_j-\textbf{r}_i
)}{|\textbf{r}_j-\textbf{r}_i|^3}- \frac{\textbf{r}_j}{r_j^3}\right]
\nonumber\\
+\textbf{F}_{\rm damp}+\textbf{F}_{\rm migI},~~~~~~~~~~~~~~~ \label{eqf}
\end{eqnarray}
where

\begin {eqnarray}
\begin{array}{lll}
\textbf{F}_{\rm damp} = -2\frac{\displaystyle (\textbf{V}_i \cdot
\textbf{r}_i)\textbf{r}_i}{\displaystyle r_i^2\tau_{\rm damp}},
\\
\cr\noalign{\vskip 0.5 mm} \textbf{F}_{\rm
migI}=-\frac{\displaystyle \textbf{V}_i}{\displaystyle 2\tau_{\rm
migI}}, \label{dm}
\end{array}
\end{eqnarray}
where $\textbf{r}_i$ and $\textbf{V}_i$ display the position and
velocity vectors of the planetary embryos. All vectors in the equations (\ref{eqf}) and (\ref{dm})  are expressed in stellar-centric coordinates.

We integrate equation (\ref{eqf}) to investigate the dynamical evolution of planetary embryos in the system using the time-symmetric integrator Hermit scheme \citep{Aarseth}. Initially, we assume that there is a solar-like central star in the system, which hosts three planetary embryos surrounded by gas and solid disks. In our numerical simulations, all planetary embryos are assumed to occupy coplanar and near-circular orbits initially. The mean anomaly, argument of pericenter, and the longitude of ascending node are generated between $0^{0}$ to $360^{0}$ randomly.

\section {Numerical Simulations and Results}
\subsection{Model Case}
Based on our previous works on formation of near 4:2:1 MMRs \citep{Wang12, Wang14}, we found that planets in the system can be captured into the configuration of near 4:2:1 MMRs under type I orbital migration. It depends on the star properties and the speed of type I migration. In this work, we generate a typical case of simulation as a model case. Three planets in the system, where P1 denotes the innermost one, P2 is the middle one, and P3 means the outermost one, can form 4:2:1 MMRs configuration. In this case, the mass of the central star is 1 $M_\odot$. The masses of three planets are 5, 10, and 15 $M_\oplus$, respectively. The magnetic field of the star is 0.5 KG, and the mass accretion rate of the star is $1.0 \times 10^{-9}~M_\odot~\rm yr^{-1}$. The timescale of type I migration we used in this case is shown as in Equation (\ref{tauI}) and $f_1=0.3$. At the beginning of the simulation, three planets locates at the orbital periods of 100, 250, and 600 days, respectively. Figure \ref{modelc} shows the evolution of orbital period in panel (a), the eccentricity variation in panel (b), the evolution of period ratios and resonance angles of each pairs of planets in panel (c), (e) and (f). Panel (d) displays the gas density profile of the disk in the simulation. In this case, the first two planets are captured into 2:1 MMR at $\sim$ 0.1 Myrs whereas P2 and P3 are trapped into 2:1 MMR at $\sim$ 0.3 Myrs. After both planets are locked into MMRs, their eccentricities are excited. The eccentricity of the first planet P1 can be pumped up to near 0.35. Moreover, due to the strong gas damping effect, then the eccentricities are damped to $\sim$ 0.01. Finally, P3 ceases at the position of the density maximum as shown in panel (d). The simulation results show that three planets locate at 18.19, 36.65, and 74.47 days, respectively. In the model case, there is no mass accretion in the formation. Three planets are assumed to be the mass at the beginning of the simulation. We use the parameters in this model case to examine the role of mass growing process.

\begin{figure*}[ht!]
\centering

  \epsscale{1.15}\plotone{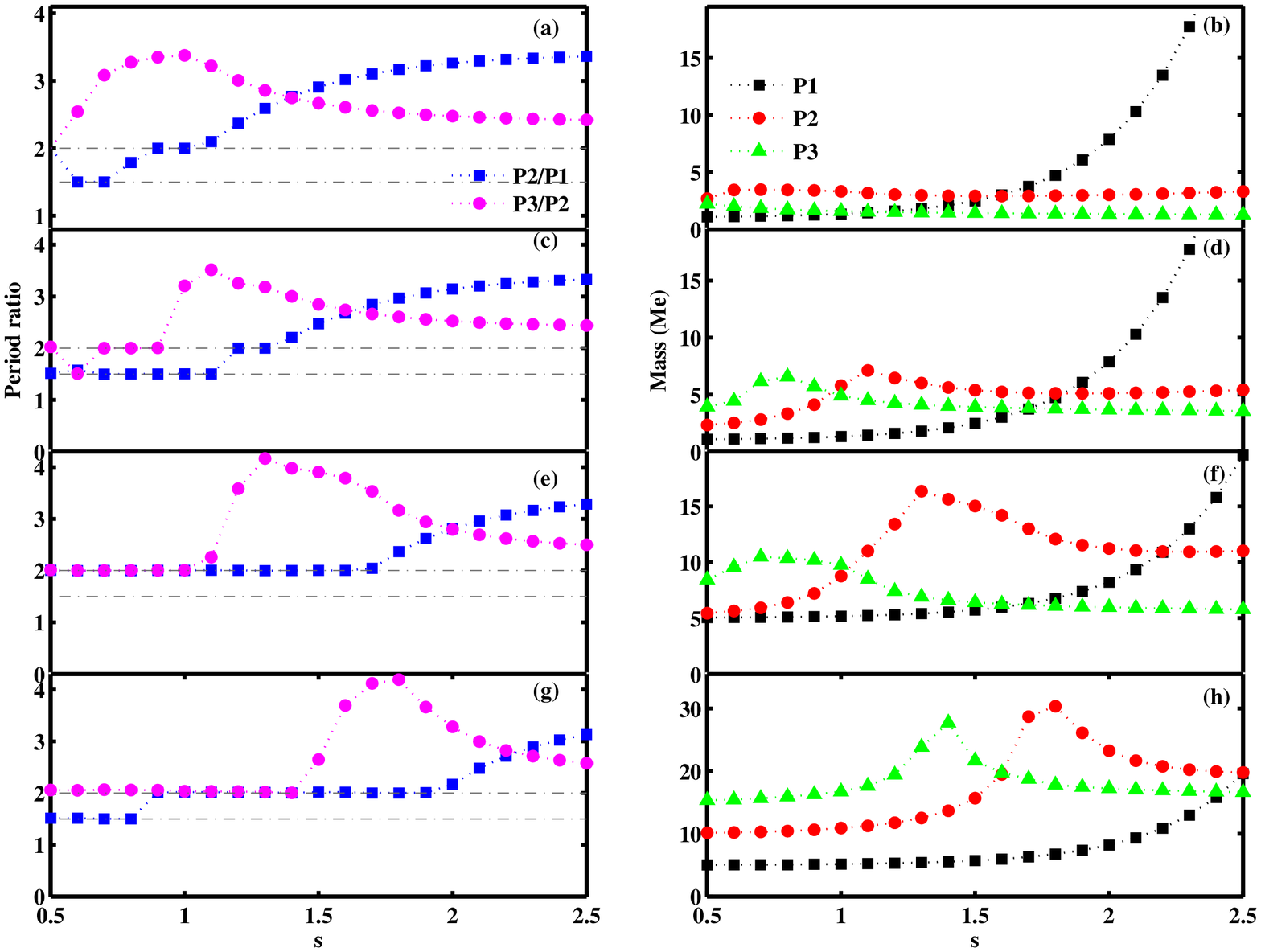}
 \caption{The results of Group 1, 2, 3, and 4. Panel (a) and (b) show the distribution of final period ratios and final masses of three planets in Group 1. Panel (c) and (d) shows the results of that in Group 2. Panel (e) and (f) shows the results of that in Group 3. And Panel (g) and (h) shows the results of that in Group 4. In Panel (a), (c), (e), and (g), the purple means the outer pair and the blue displays the inner pair. In panel (b), (d), (f), and (h), the black squares show the masses of P1, the red dots mean the masses of P2, and the green triangles represent the masses of P3.
 \label{all1}}
\end{figure*}

\begin{table}
\centering \caption{The initial masses of three planets in the system and the power law index of the gas disk density in G1, G2, G3 and G4.
 \label{tb1}}
\begin{tabular*}{8cm}{@{\extracolsep{\fill}}ccccc}
\tableline
 & $m_{10}$& $m_{20}$&$m_{30}$& $s$\\
 &$(M_\oplus$) &($M_\oplus$) &($M_\oplus$)&  \\
\tableline
G1&1 &1 &1 & [0.5 2.5]\\
G2&1&2&3 &[0.5 2.5]\\
G3&5 &5 &5 & [0.5 2.5]\\
G4&5&10&15 &[0.5 2.5]\\
G5&1&1&1&[0.5 2.5]\\
G6&1&isolation mass&isolation mass&[0.5 2.5]\\
\tableline
\label{group}
\end{tabular*}
\end{table}

\subsection{The cases with mass accretion process and inward type I migration}
We use all initial parameters as given in the model case except the masses of three planets. The initial masses of the planets are shown in Table \ref{group}. Considering various initial masses, we perform four groups of simulations. In the investigation of four groups, we take into account the value of $s$ in the range of [0.5, 2.5] in an aim to examine the role of different profiles of gas and solid density.

Group 1: the initial mass of three planets are all set to be 1 $M_\oplus$.

Group 2: According to the estimation of isolation mass \citep{idalin04}, the masses of planets are proportion  to $a^{(3-3s/2)}$. This means that when $s$ is less than 2, the masses of planets increase along with the semi-major axis. Therefore, in our simulations, we set the initial masses of planets to be $m_1=1~M_\oplus$, $m_2=2~M_\oplus$, and $m_3=3~M_\oplus$, respectively.

Group 3: we remain the initial masses of planets to be a higher value, which are set to be $m_1=m_2=m_3=5~M_\oplus$.

Group 4: we set the initial masses of planets to be $m_1=5~M_\oplus$, $m_2=10~M_\oplus$, and $m_3=15~M_\oplus$, which is identical to the model case. But in this Group, the masses of planets can increase as time evovles.

According to the value of $s$, there are 21 runs for each Group. The major initial conditions are listed in Table \ref{tb1}, and the leading outcomes are shown in Figure \ref{all1}.

\subsubsection{G1: The trapping of near mean motion resonances under inward migration with low equal masses}

 In this group, the initial masses of three planets are 1 $M_\oplus$. Panel (a) and (b) in Figure \ref{all1} shows the results of Group 1. Panel (a) shows the distribution of the final period ratios. The purple ones represent the period ratio between P3 and P2 while the blue squares display the period ratio between P2 and P1. From the distribution in panel (a), we can find that the outer pair can be trapped into 2:1 MMRs only when $s=0.5$. The reason leading to such low possibility to be in 2:1 MMR is the mass relationship between P3 and P2. As shown in panel (b), the mass of P2 is always more massive than P3. According to the timescale of type I migration in Equation (\ref{tauI}), high mass planets with smaller semi-major axis will be involve in a shorter timescale. Then the outer small planet cannot catch up with the middle planet P2. Therefore, the outer pair is difficult to generate the MMR configuration. However, the inner pair can enter into a 3:2 MMR when $s\in [0.6~0.7]$ and in four cases they can be captured into 2:1 MMRs for $s=0.5$ and $s\in[0.9, 1.1]$.

Panel (b) shows the distribution of the final masses of three planets. The masses of P1, P2, and P3 are represented by the color black, red, and green, respectively. In this group, the mass of P1 increases as $s$ increases. The masses of the middle planet P2 changes slightly. It is almost flat at $\sim$ 3.0 $M_\oplus$. The mass of P1 ranges in [1.1, 30.5] $M_\oplus$, whereas the mass of P3 fluctuates from 1.2 to 2.2 $M_\oplus$. For $s\in [1.6, 2.5]$, the mass of P1 is the highest among them. The detail of their mass relationship are shown in Table \ref{tb2}.

As shown in Figure (\ref{all1}),when the planets undergo inward type I migration, the planet pairs are easily to be trapped into MMRs when the gas disk is relatively flat. The evolution process is similar to the pattern shown in Figure (\ref{modelc}). The possibilities of planets trapped in 2:1 and 3:2 are $\sim$ 11.9\% and 4.8\%, respectively. Only one run in the simulation can produce the configuration of near 4:2:1 MMRs.

\begin{table*}
\centering \caption{The relationship of final masses of planets.
 \label{tb2}}
\begin{tabular*}{14cm}{@{\extracolsep{\fill}}cccccc}
\tableline
 & $m_3>m_2>m_1$&$m_2>m_3>m_1$& $m_2>m_1>m_3$& $m_1>m_2>m_3$\\
\tableline
G1&&[0.5 1.1]& [1.2 1.5]& [1.6 2.5]\\
G2&[0.5 0.9]&[1.0 1.7]&1.8&[1.9 2.5]\\
G3&[0.5 1.0]&[1.1 1.6] &[1.7 2.2]&[2.3 2.5]\\
G4&[0.5 1.6]&[1.7 2.4]&2.5 &\\
G5&&[0.9 1.5]&[1.6 2.3]&[0.5 0.8][2.4 2.5]\\
G6&[0.5 1.6]&[1.7 1.8]&[1.9 2.3]&[2.4 2.5]\\

\tableline
\label{mas}
\vspace{1.0cm}
\end{tabular*}
\end{table*}

\begin{figure*}[ht!]
\begin{center}

  \epsscale{1.15}
  \plotone{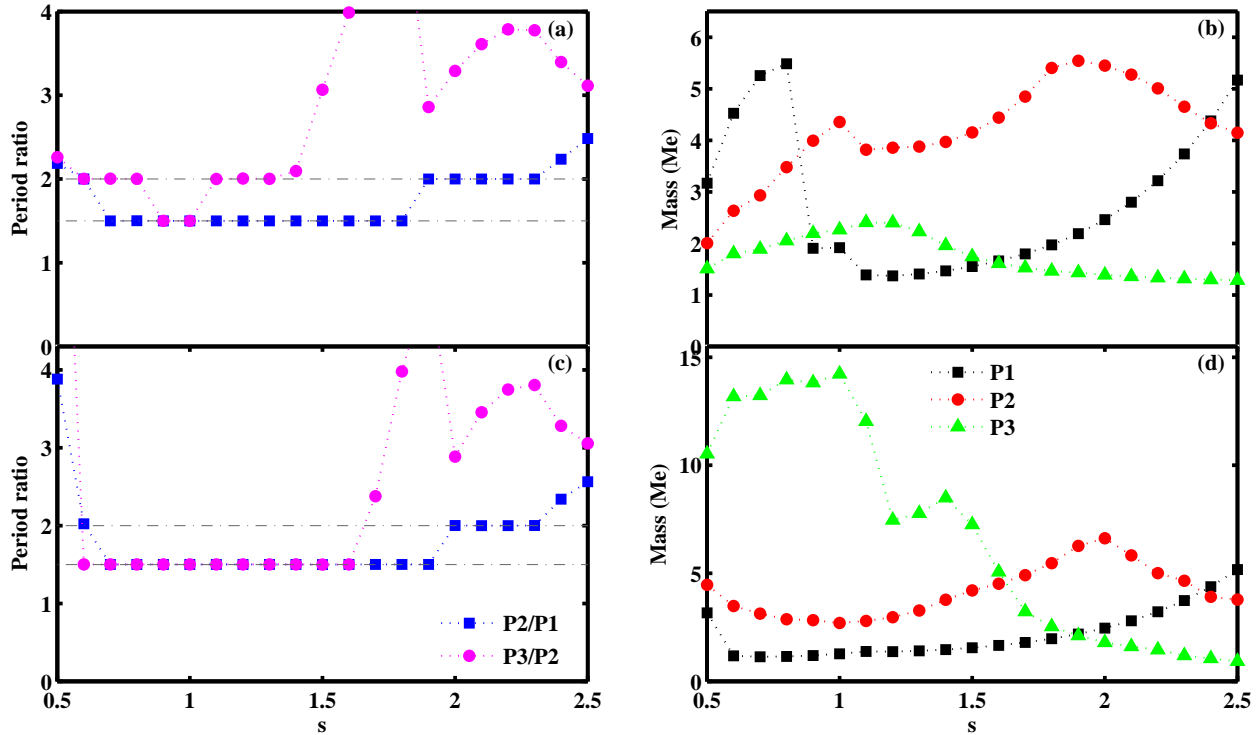}
   \vspace{-2cm}
 \caption{The results of Group 5 and 6. Panel (a) and (b) show the distribution of final period ratios and final masses of three planets in Group 5. Panel (c) and (d) shows the results of that in Group 6. In Panel (a) and (c), the purple means the outer pair and the blue displays the inner pair. In Panel (b) and (d), the black squares show the masses of P1, the red dots mean the masses of P2, and the green triangles represent the masses of P3.
 \label{outward}}
 \end{center}
\end{figure*}

\begin{figure*}
\begin{center}

  \epsscale{1.15}\plotone{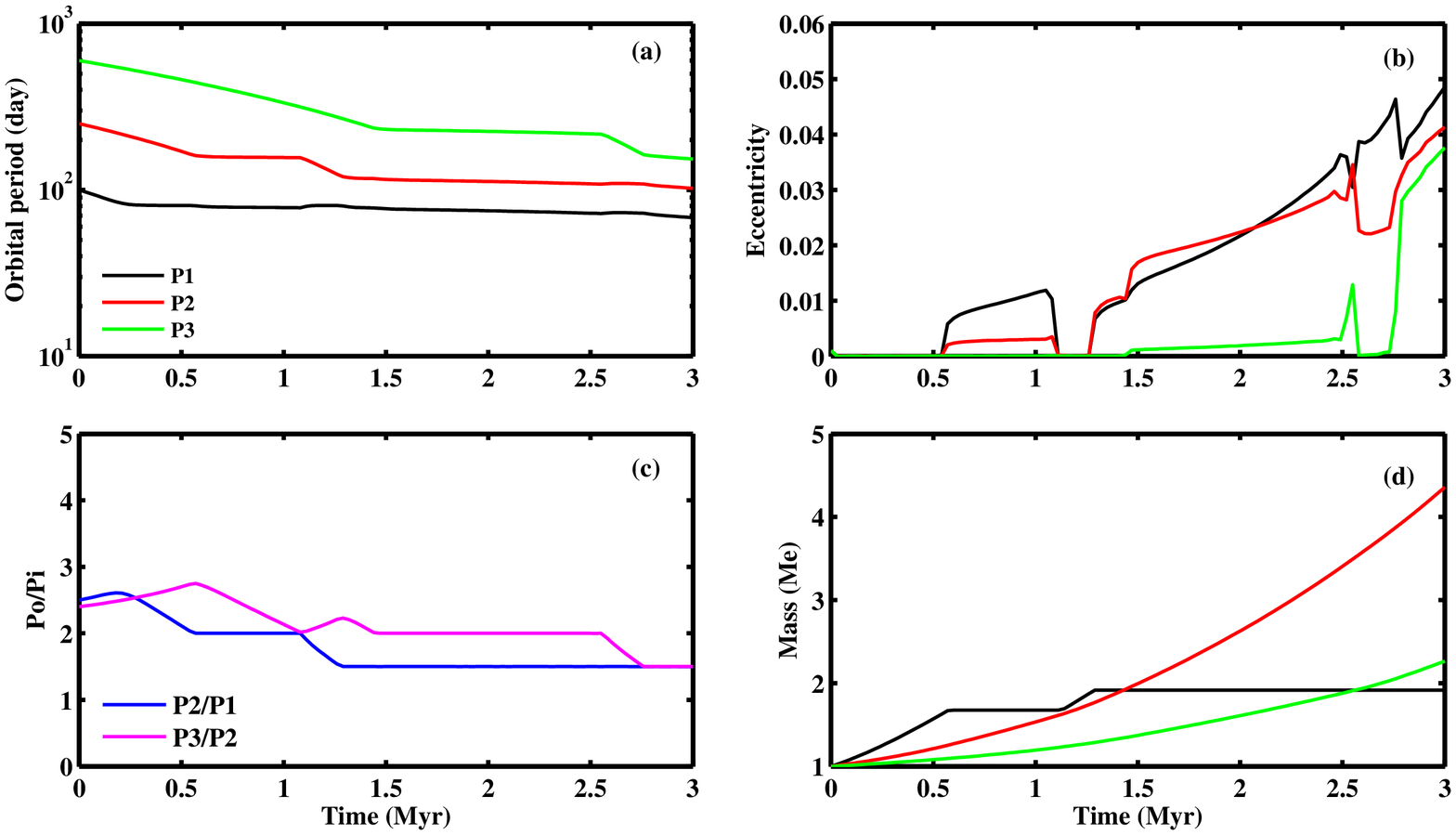}
\vspace{-2cm}
 \caption{One typical results in Group 5 with $s=1.0$. Panel (a), (b), (c) and (d) show the evolution of orbital periods, eccentricities, period ratios, and the masses of three planets. The black, red, and green lines represent the innermost planet, the middle one, and the outermost planet, respectively. In panel (c), the blue line shows the evolution of P2/P1, and pink line means the evolution of P3/P2.
 \label{single}}
 \end{center}
\end{figure*}

\subsubsection{G2: The trapping of near mean motion resonances under inward migration with 1, 2, and 3 $M_\oplus$}

In this Group, the initial masses of three planets are 1, 2, and 3 $M_\oplus$, respectively.  From the period ratio distribution shown in Panel (c) of Figure \ref{all1}, we can observe that more pairs of planets can be trapped into 3:2 MMRs than those of Group 1. When $s\leq 1.3$, at least one planet pair can be locked into MMRs. Herein the possibility that planet pairs moving into 3:2 MMRs is 19.0\%, whereas that of 2:1 MMRs is 14.3\%. For $s=0.6$, two planets are both in 3:2 MMRs, leading to chain resonances of 9:6:4, whereas for $s=0.5$ and $s \in [0.7, 0.9]$, three planets are in the 6:3:2 MMRs configuration. The relationship of the masses of three planets follows $m_3>m_2>m_1$ when the system can form chain resonances configuration.

From Panel (d) of Figure \ref{all1}, we can see that the distribution trend of P1 is similar to that of Group 1. The mass of P1 is in the range [1.07, 30.5]. The mass of P2 peaks at $s=1.1$ with $m_2=7.1$ $M_\oplus$ and the mass of P3 peaks at $s=0.8$ with $m_3=6.6$ $M_\oplus$. The mass fluctuation of P2 and P3 are small. Most of their masses are lower than 5 $M_\oplus$.  In addition, $m_2$ is equal to $m_3$ in the case so that $s$ meets $0.9<s=s0<1.0$, whereas three planets bear similar masses for $1.7<s=s1<1.9$.  When $s1>s>s0$, the mass of the middle planet is massive that the other two, and when $s<s0$, the innermost planet is the massive one. The results show that the possibility of capture into MMRs for the planet pairs decreases as the mass of P1 increases in the evolution.

\subsubsection{G3: The trapping of near mean motion resonances with massive equal masses}

In this Group, each of the initial masses of three planets is adopted to be 5 $M_\oplus$.  As shown in Panel (e) of Figure \ref{all1}, there is no 3:2 MMRs survival in the simulations.  However, 19 planet pairs are found to be trapped in 2:1 MMRs, occupying a fraction of 45.2\% among all pairs. Furthermore, we find that the fraction of two inner planets and the outer pair is 30.9\% and 14.3\% among all pairs, respectively, indicating that the inner pair is more easily to be in MMRs than the outer two companions.

From Panel (f) of Figure \ref{all1}, we can get that P3 is the massive one of them and both two pairs will be in 2:1 MMRs when $s\leq1.0$ . Thus, when the outermost planet is more massive than the other two planets, they are easily to be in 4:2:1 MMRs. According to the definition of isolation mass \citep{idalin04}, such mass configuration is natural to form with MMSN.  When $s>1.7$, the planet pairs are difficult to be in MMRs, regardless of 2:1 or 3:2 MMRs. In a word, the distribution of planet masses are similar to that of Group 2, except that the mass of the middle planet has a larger range. The mass of P2 can reach 16.4 $M_\oplus$.

\subsubsection{G4: The trapping of near mean motion resonances with same masses as in the model case initially}

In this Group, the initial masses of planets are 5, 10, 15 $M_\oplus$ respectively, which are the same as in the model case. Similar to that in Group 3, plenty of planet pairs are found to be trapped into MMRs eventually, e.g, 2:1 MMRs. In the simulations, there are six runs that both two pairs are in 2:1 MMRs. The 4:2:1 MMRs configuration is similar to that in the model case. Besides, there are nine systems with one planet pair in 2:1 MMRs. In summary, almost 50\% of all pairs are locked into 2:1 MMRs. But there are four pairs in 3:2 MMRs only with a possibility of 9.5\%. When $s>1.9$, the planets have a difficulty in evolving into MMRs. As previously reported, this possibility of planets in MMRs is similar to that in the model case \citep{Wang14}.

As shown in Panel (h) of Figure \ref{all1}, the final masses of planets can reach $\sim$ 30 $M_\oplus$. Increased with $s$, the mass of planet 1 increases, and finally reach almost 20 $M_\oplus$. The mass of P2 peaks at $s=1.8$ with $\sim$ 30 $M_\oplus$, whereas the mass of P3 peaks at $s=1.4$ with $\sim$ 30 $M_\oplus$. The mass of P3 ranges in the [15.4, 27.7] $M_\oplus$, which can be comparable to the mass of P2. P1 always has the lowest mass. When $s>1.6$, the middle planet has the highest mass. For $s>1.9$, the masses of three planets are close to each other. Due to a steeper profile of disk density, they are hard to be in MMRs.

From the simulations of above-mentioned four Groups, we can conclude that if the planetary masses increase with the semi-major axis in one system, the possibility of planet pairs in that system trapped into MMRs is high. Planet pairs holds the highest possibilities in Group 4, while the fraction is the lowest in Group 1. In Group 2, the probability of planet pairs in 3:2 MMRs is the highest. The mass ratio of the planet pairs in the range of [2.1, 5] when they are in 3:2 MMRs. By investigating all cases of four Groups, we can learn that there are 28.6 \% pairs in 2:1 MMRs and 8.3\% in 3:2 MMRs.

\subsection{The cases with mass accretion process and possible outward type I migration}
 In our simulations, the masses of planets are around ten to a couple of earth-mass, the orbital migration is mainly triggered by the Lindblad and Corotation torque. The timescale of type I migration in equation (\ref{tauI}) which leads to an inward migration is derived from linear analysis of isothermal gaseous disk. But the corotation torques are generally non-linear. Considering the Lindblad torque and non-linear horseshoe drag in adiabatic model, the timescale of type I migration can be expressed as \citep{Paa10, Paa11}
\begin{eqnarray}
\tau_{\rm migI1}=\frac{a}{\dot{a}}=\frac{m\sqrt{GM_*a}}{a\Gamma_{\rm total}},~~~~~~~~~~~~~~~~~~~~~~~~~~~~~~~
\nonumber\\
\Gamma_{\rm total}=\Gamma_{\rm L}+\Gamma_{\rm c,hs}~~~~~~~~~~~~~~~~~~~~~~~~~~~~~~~~~~~~~~~~
\nonumber\\
=\frac{\Gamma_0}{\gamma}[(-2.5-1.7b+0.1s)(\frac{0.4}{c/h})^{0.71}~~~~~~~~~~
\nonumber\\
+\frac{0.44}{c/h}(1.5-s)+\frac{\xi}{\gamma}\frac{0.4}{c/h}(10.1\sqrt{\frac{0.4}{c/h}}-2.2)]
\label{ttorq}
\end{eqnarray}
where $\xi=b-(\gamma-1)s$, $\Gamma_0=(q/h)^2\Sigma_ga^4\Omega^2$ and $\Gamma_{\rm total}$ is the total torque caused by the Lindblad ($\Gamma_{\rm L}$) and horseshoe drag ($\Gamma_{\rm c,hs}$). $b$ and $\gamma$ are the power law index of the temperature profile and the adiabatic index, respectively. $q=m/M_*$ is the mass ratio between the planetary embryo and the central star.

According to this kind of migration, planet will suffer from outward migration when the total torque is positive. Therefore, the migration mode will have significant influence on the final configuration of the planetary system. In order to understand the role of the outward migration, we carry out additional  simulations for Group 5 and Group 6, in which different initial masses of planets are set under consideration of orbital migration in Equation (\ref{ttorq}).

Group 5: The initial masses of three planets are all set to be 1 $M_\oplus$, as indicated in Group 1.

Group 6: The initial mass of P1 is 1 $M_\oplus$, while those of P2 and P3 are estimated from the isolation mass \citep{idalin04}.

Figure \ref{outward} shows the major statistics of distribution of final period ratios and final masses for three planets.

\subsubsection{G5: The trapping of near mean motion resonances with small equal masses}

In Group 5, as aforementioned, the initial masses of three planets start from 1 $M_\oplus$, and they can accumulate by accreting  the material from the disk in the evolution. The migration timescale can be estimated from Equation (\ref{ttorq}). The final distribution of the orbital period ratio and final masses are shown in Panel (a) and (b) of Figure \ref{outward}. From Panel (a), we can note that there are 14 planet pairs trapped in 3:2 MMRs and 12 planet pairs in 2:1 MMRs. The probability of planet pairs in 3:2 MMRs is $\sim$ 33.3\%, whereas that of two planets trapped in 2:1 MMRs is $\sim$ 30.9\%. Interestingly, two runs show that both of two pairs are in 3:2 MMRs in each system. However, when $0.7\leq s\leq1.8$, the occurrence of planet pair in 3:2 MMRs appears to be significantly high.

From Panel (b), we can find that the masses of three planets fluctuate within a large amplitude. When $s<0.9$, P1 owns the most material of the system. Otherwise, the middle planet is the massive one in the system. When $0.9\leq s \leq1.8$, planet 1 and 3 contain comparable mass and the two inner planets are always captured into 3:2 MMRs in this range. The final masses of three planets are in the range of [1.1, 5.8] $M_\oplus$. Figure \ref{single} shows one of the typical simulation results in Group 5. Panel (a), (b), (c) and (d) exhibit the evolution of orbital periods, eccentricities, period ratios, and the masses of three planets. From Figure \ref{single}, we can safely draw a conclusion that the two inner planets can be trapped into 2:1 MMRs when the first planet ceases at the boundary of the inner hole. However, when the third planet moves, it can drive the middle planet migrate inward and expel the innermost planet , which migrates outward. Finally, the inner pair escapes from 2:1 MMR to 3:2 MMR at $\sim$ 1.2 Myr, which is caused by the outward migration of the innermost planet at the time. The outer pair will depart from 2:1 MMR to 3:2 MMR at the end of evolution due to the mass accretion scenario. The masses of planets has been increased to be 1.9, 4.4, and 2.2 $M_\oplus$, with the orbital period of 68, 102, and 153 days, respectively.

\subsubsection{G6: The trapping of near mean motion resonances with isolation masses}

As shown in Panel (c) of Figure \ref{outward}, we can observe that two pairs are both trapped into 3:2 MMRs of 10 runs in Group 6. Moreover, there are three runs that the inner pair is in 3:2 MMRs, and one case with the two outer planets in 3:2 MMR. The probability that the planet pair trapped in 3:2 MMRs is $\sim$ 57.1\%. It is significantly higher than that in other Groups. The possibility that planet pairs captured into 2:1 MMRs is lower than that of 3:2 MMRs. There are five runs with the inner pair in 2:1 MMRs, $\sim$ 11.9 \%.

The final mass of P3 decreases along with the increase of $s$, and when $s\in [0.5, 1.6]$, P3 is the most massive planets in the system with respect to $m_3>m_2>m_1$. The mass of P1 increases as $s$ increases, and $m_1$ is more massive than $m_3$ for $s>1.9$. Herein the masses of P1 and P2 range in the [1.1, 6.6] $M_\oplus$, whereas the mass of P3 is in a wider range of [1.1, 14.2] $M_\oplus$.

From the results of Group 5 and 6, we can conclude that the likelihood of planet pairs in 3:2 MMRs can significantly rise. In addition to the mass accretion, the outward migration occurred in the inner region plays a vital role in forming final configuration of the system during this process. Considering outward migration, the mass of the innermost planet cannot grow much larger, implying that an alternative fact leading to a high possibility of 3:2 MMRs. Combined with all runs in Group 5 and 6, we find that there are 20.2\% pairs in 2:1 MMRs and 45.2\% in 3:2 MMRs. Taking into account all cases in Group 1-6, we have performed 126 simulations in total, among which 68 planet pairs are found in 2:1 MMRs and 52 pairs in 3:2 MMRs in the final configuration. From our simulation outcomes, we further find many chain resonance systems that eventually formed, among which 14 systems are in 4:2:1 MMRs, 14 systems in 6:3:2 MMRs, 13 systems in 9:6:4, and 1 system in 3:2:1 MMRs.  These simulations suggest that our formation scenario can be used to explain some chain resonances formation in the planetary systems.

\section{Conclusions and Discussions}

In this work, we mainly investigated the formation of planet pairs in MMRs by considering the scenario of the mass accumulating and potential outward orbital migration. We performed 126 runs of simulations with various initial masses of planets and migration modes. From the simulations, we find that the mass accretion can change the final mass distribution of planets, thereby altering the final configuration of systems. The outward migration plays a positive part in increasing the likelihood of planet pairs trapped into 3:2 MMRs.

\begin{enumerate}
\item
Mass accretion scenario in our simulations can make the planets grow from 1 $M_\oplus$ to $\sim$ 30 $M_\oplus$.  And the mass variation during the formation process can boost the opportunity of planets captured into 3:2 MMRs, especially with the mass ratio between [1.5, 5.0]. By examining the results of six groups, we find that there exists a high likelihood for the occurrence of 2:1 MMRs regardless the initial masses, and 3:2 MMRs are more likely to happen in the inner pair than in the outer pair. We point out that most of 2:1 and 3:2 MMRs took place in the system with relatively flat disks, mainly in the range of $s<2.0$. In this range, usually the innermost planet has a lowest mass in the system. The systems whose planetary masses increase with the distances from the central star, have more chance to yield 4:2:1 MMRs configuration. According to the estimation of isolation mass \citep{idalin04}, such mass relationship is easy to form. Consequently, the near 4:2:1 MMRs appears to be common in planetary systems \citep{Wang12, Wang14, Sun17}. From the statistics of mass relationship among three planet in Table \ref{mas}, we find that the possibility of planet pairs in MMRs is very low when the mass of the outmost planet is the smallest one in the system. Therefore, under an inward migration, the mass relationship of three planets acts as a major factor leading to the capture of MMRs and the 3:2 MMRs can further be enhanced in the mass growing process.

\item
If the planets can undergo outward migration during the formation process, the possibility of planet pairs in 3:2 MMRs will be increased significantly. The final masses of planets can grow up to $\sim$ 15 $M_\oplus$, which is lower than that in the case with only inward migration. With a potential outward migration, both of the inner planet pair and the outer planet pair can be easily trapped into 3:2 MMRs. The planet pairs have opportunities to be in MMRs even in disks with high $s$ which represents a steep profile of density. From Table \ref{mas}, we can show that if the mass of the outermost is much higher than that of the other two planets, both planet pairs can be captured into 3:2 MMRs easily. Hence, the outward orbital migration can play an efficient part in resulting in the capture of 3:2 MMRs.
\end{enumerate}

Based on our results, the mass accumulation and the direction of type I orbital migration are two crucial factors to drive the planet pairs moving into 3:2 MMRs, whereas 2:1 MMRs is an evolution outcome, which makes the Laplace Resonance occur as observed. Overall, there are 25.8\% planet pairs in 2:1 MMRs and 20.6\% planet pairs in 3:2 MMRs.

\begin{figure}
\begin{center}
  \epsscale{1.2}\plotone{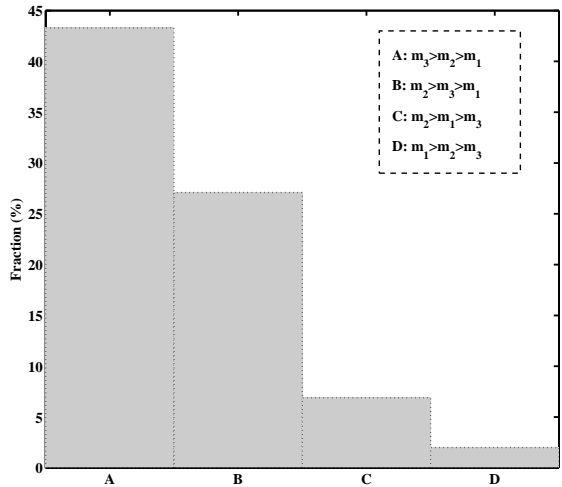}

 \caption{The mass relationship obtained from the observation data in three-planet system. A, B, C, and D display four different mass relationship among planets
 \label{ob}}
 \end{center}
\end{figure}

\begin{table*}
\centering \caption{The orbital periods and planet radii in seven near chain resonances systems.
\label{result}}
\begin{tabular*}{16cm}{@{\extracolsep{\fill}}lccccccc}
\tableline
\hline
Name&& & Orbital Period\footnote{https://exoplanetarchive.ipac.caltech.edu} &Planet Radii& Radii (+)& Radii(-)& Star Mass\\
&&&(day)& ($R_\oplus$) & ($R_\oplus$)& ($R_\oplus$)&($M_\odot$)\\
\tableline
\hline
 &d&K00829.02&9.75& 2.44& 0.94 & -0.23&\\
Kepler-53& b&K00829.01&18.65& 2.90& 1.12 & -0.27&1.019\\
 &c&K00829.03&38.56& 3.57& 1.37 & -0.33&\\
\tableline

 &b&K02707.02&14.43& 1.01& 0.18 & -0.07&\\
Kepler-399& c&K02707.01&26.68& 1.58& 0.28 & -0.11&0.707\\
 &d&K02707.03&58.03& 1.96& 0.36 & -0.14&\\
\tableline
 &b&K01426.01&38.87& 3.31& 1.52 & -0.33&\\
Kepler-297& c&K01426.02&74.93& 7.54& 3.47 & -0.75&1.040\\
  &d &K01426.03 &150.02& 13.54& 6.22 & -1.35&\\
 \tableline
  &d&K00279.03&7.51& 0.89& 0.05 & -0.07&\\
Kepler-450& c&K00279.02&15.41& 2.49& 0.13 & -0.14&1.350\\
 &b&K00279.01&28.45& 5.93& 0.76 & -0.18&\\
 \tableline
 \hline
   &d&K00584.03&6.47& 0.89& 0.29 & -0.17&\\
Kepler-192& b&K00584.01&9.93& 2.20& 0.74 & -0.17&0.891\\
 &c&K00584.02&21.22& 2.27& 0.32 & -0.32&\\
 \tableline
 \hline
    &b&K00148.01&4.78& 1.89& 0.23 & -0.06&\\
Kepler-48& c&K00184.02&9.67& 2.61& 0.32 & -0.09&0.955\\
 &d&K00184.03&42.90& 2.02& 0.25 & -0.07&\\
 \tableline
      &b&K00041.02&6.89& 1.31& 0.03 & -0.03&\\
Kepler-100& c&K00041.01&12.82& 2.28& 0.06 & -0.05&1.080\\
 &d&K00041.03&35.33& 1.50& 0.03 & -0.04&\\
 \tableline
 \hline
\vspace{1.0cm}
\end{tabular*}
\end{table*}

Using the released data of Kepler Mission, we carry out a statistical analysis of the mass relationship in three-planet systems. The results are shown in Figure \ref{ob}. Herein 203 three-planet systems are included. A, B, C, D, which stand for four various mass relationship of the planets, i.e., $m_3>m_2>m_1$, $m_2>m_3>m_1$, $m_2>m_1>m_3$ and $m_1>m_2>m_3$, respectively. We assume all planets have similar densities. From Figure \ref{ob}, we can find that the relationship A is the most popular case in planetary systems, indicating that these systems may possess planets in MMRs according to our simulations. More than 40\% of systems follow such mass relationship. If there exists MMRs in the systems, the planets are formed in relative flat disks with $s<1.5$. Next, the mass relationship B and C suggest that the middle planet is more massive than the other two companions in the same planetary system, thereby the inner planet pair is easily involved in MMRs in such system. And the system may be evolved from a disk with steep distribution. The relationship D with $m_3<m_2<m_1$ could occupy the lowest fraction among the above-mentioned cases of mass relationship. In comparison with our simulations, there are several cases leading to such mass relationship when $s$ is larger than 2 or has a much larger value. Therefore, the system, which hosts an innermost planet with the highest mass among three planets, may be produced in an extremely steep profile of disks. But under the effect of such disks, the planets are difficult to be trapped in MMRs. In summary, we can draw a safe conclusion that the MMRs configurations can be easily produced for the systems by Kepler mission when considering a higher fraction of A and B.

Our simulations further show that the chain resonances, e.g., 4:2:1, 3:2:1, 6:3:2 and 9:6:4 MMRs, have been revealed in the systems. The recent observations suggest that the chain resonances appear to be common in the planetary systems. Our formation scenarios proposed in this work can be applicable to the chain resonances formation for the systems like K2-72 \citep{WM14, Cre16, Dr17} and TRAPPIST-1\citep{Gi17}. Furthermore, more examples are taken from the three-planet systems confirmed by Kepler mission, and we find that seven systems (see Table \ref{result}), whose orbital configurations are in near chain resonances, can be also explained by our formation scenario. As summarized in Table \ref{result}, Kepler-53, Kepler-399, Kepler-297 and Kepler-450 are in near 4:2:1 chain resonances, among which two planet pairs are close to 2:1 MMRs. And their planetary masses follow the relationship A : $m_3>m_2>m_1$. Three planets in Kepler-192 are in the near 6:3:2 MMRs, where the middle planet and the outermost planet are much larger than the innermost one. Such mass relationship and orbital configuration can be formed through the formation scenario proposed in this work. The mass relationship for Kepler-48 and Kepler-100 which host two inner planets in near 2:1 MMRs, is consistent with relationship B: $m_2>m_3>m_1$. Such systems may be formed in the disk with a high $s$ as shown in Figure \ref{all1} and \ref{outward}.

In this work, we simply consider the formation process from planetary embryos to the evolution stage when they enter into MMRs. In this model, the mass accretion may alter the orbits of planets and affect the efficiency of orbital migration. Because the timescale of mass accretion in this work is much longer than that of type I migration, thus we did not include the possible orbital variations caused by the mass accretion in our model. From the statistical results of Kepler data \citep{lee13, Wang14}, we can observe that most planet pairs are not in the exact location of MMRs. The mechanism of driving the planet pairs deviate from MMRs to near MMRs plays a crucial role of understanding the formation of systems. For those systems bearing planets with close-in orbits, tidal dissipation arising from the central star will be at work, thereby making the planets detached from the resonances \citep{ML04, ZL08, LW12, BM13, Dong13, lee13, Sun16, Dong17}.  When the planets are far from the central star, the tidal dissipation is not strong enough, the depletion timescale of gas disk will play an important role in making them moving out of resonances (Wang et al. in prep.).

\acknowledgments{This work is supported by National Natural
Science Foundation of China (Grants No. 11573073, 11773081, 11473073, 11661161013), the Natural Science Foundation of Jiangsu Province (Grant No. BK20151607, BK20141509), the Strategic Priority Research Program-The Emergence of Cosmological Structures of the Chinese Academy of Sciences (Grant No. XDB09000000), the innovative and interdisciplinary program by CAS (Grant No. KJZD-EW-Z001), the Foundation of Minor Planets of Purple Mountain Observatory, and Youth Innovation Promotion Association.}

\end{document}